\documentclass[twocolumn]{aastex701}

\usepackage{amsmath}
\usepackage{amssymb}
\usepackage{graphicx}
\usepackage{booktabs}
\usepackage{iftex}
\ifXeTeX
  \usepackage{xeCJK}
  \newcommand{\CJKname}[1]{#1}
\else
  \usepackage{CJKutf8}
  \newcommand{\CJKname}[1]{\begin{CJK*}{UTF8}{gbsn}#1\end{CJK*}}
\fi
\newcommand{\Gaia}{\textit{Gaia}}
\newcommand{\code}[1]{\texttt{#1}}
\newcommand{\Aseven}{\code{Acceleration7}}
\newcommand{\Anine}{\code{Acceleration9}}
\newcommand{\masyrtwo}{mas yr$^{-2}$}

\newcommand{\lsun}{\ensuremath{L_\odot}}
\newcommand{\pirs}{\pi_{\rm res}}

\begin{document}
\raggedbottom
\shorttitle{Gaia stellar engine technosignatures}
\shortauthors{Huang et al.}

\title{Gaia DR3 Limits on Stellar Engine Technosignatures in Nearby Stars}

\correspondingauthor{Tong-Jie Zhang}
\email{tjzhang@bnu.edu.cn}

\author[orcid=0000-0002-8719-3137,sname='Huang']{Bo-Lun Huang (\CJKname{黄博伦})}
\affiliation{Institute for Frontiers in Astronomy and Astrophysics, Beijing Normal University, Beijing 102206, China}
\affiliation{School of Physics and Astronomy, Beijing Normal University, Beijing 100875, China}
\email{Bolunh@hotmail.com}

\author[orcid=0000-0002-4683-5500,sname='Tao']{Zhen-Zhao Tao (\CJKname{陶振钊})}
\affiliation{College of Computer and Information Engineering, Dezhou University, Dezhou 253023, China}
\email{tzzzxc@163.com}

\author[orcid=0000-0002-3363-9965,sname='Zhang']{Tong-Jie Zhang (\CJKname{张同杰})}
\affiliation{Institute for Frontiers in Astronomy and Astrophysics, Beijing Normal University, Beijing 102206, China}
\affiliation{School of Physics and Astronomy, Beijing Normal University, Beijing 100875, China}
\email{tjzhang@bnu.edu.cn}

\author[orcid=0000-0003-3153-1296]{Zhi-E Liu (\CJKname{刘志娥})}
\affiliation{College of Physics and Electronic Engineering, Qilu Normal University, Jinan 250200, People’s Republic of China}
\email{zhieliu@qlnu.edu.cn}

\begin{abstract}
An operating stellar engine that transfers momentum to its host star could produce an approximately persistent transverse acceleration during the \Gaia\ Data Release 3 (DR3) observing epoch.
We ask how common such stellar-engine-like accelerations could be among nearby stars.
We define a quality-selected parent population of $406{,}984$ \Gaia\ DR3 stars with nominal distances below 200 pc (about 650 light-years) and analyze the $8{,}890$ published constant-acceleration (\Aseven) solutions within this population.
Because long-period binaries and astrometric systematics can produce similar sky-plane curvature, we model these conventional explanations together with a broad component that can absorb persistent acceleration or remaining model mismatch.
Above a transverse-acceleration threshold of $a_0=2.60\times10^{-4}\ {\rm m\ s^{-2}}$, where the modeled searchable fraction is essentially unity, the one-sided 95\% model-conditional upper limit on the fraction of the parent population allowed to show such accelerations is $1.21\times10^{-5}$---about 12 per million stars in this parent population.
At lower acceleration thresholds, the modeled searchable fraction declines, so the corresponding bounds mainly characterize the reach of the search.
No individual source is interpreted as a technological signal.
\end{abstract}

\section{Introduction}\label{sec:intro}

The search for extraterrestrial intelligence (SETI) has expanded rapidly in the twenty-first century, encompassing a broad range of theoretical frameworks and observational searches for technosignatures \citep{LingamLoeb2021,Wright2026}.
\hypersetup{citecolor=blue}
Most such searches analyze information carried by electromagnetic radiation, including narrowband radio emission, optical pulses, atmospheric or surface signatures, and infrared excess from large-scale energy use \citep{Tarter2001,Dyson1960,Wright2014,HaqqMisra2022,Wright2022,huang2026wise}.
Astrometry offers a complementary route: not unusual light, but unusual motion.
A technology that transfers momentum to a star could contribute a sustained astrometric acceleration while it operates.
This motivates the central question of this work: above a given acceleration threshold, what fraction of nearby stars could show a sustained transverse acceleration after allowing for ordinary binary motion and catalog systematics?

The deliberate alteration of stellar or planetary motion has a long conceptual history, from Zwicky's early proposal to the explicit stellar-engine designs of Shkadov and later authors \citep{Zwicky1957,Shkadov1987,BadescuCathcart2000,BadescuCathcart2006,LingamLoeb2020}.
Proposed mechanisms include radiation-redirecting structures, active momentum transfer, and steerable binary-engine designs \citep{Shkadov1987,BadescuCathcart2000,BadescuCathcart2006,Caplan2019,Svoronos2020,LingamLoeb2020,Vidal2024}.
One long-term motivation is control of a planetary system's Galactic trajectory, as considered in earlier stellar-engine studies \citep{Shkadov1987,BadescuCathcart2006}.
Related astrobiological work has discussed whether vertical Galactic motion, spiral-arm passages, nearby supernovae, cosmic rays, or comet injection could affect terrestrial environments over tens of Myr \citep{BahcallBahcall1985,GiesHelsel2005,RohdeMuller2005,MedvedevMelott2007,BailerJones2009}.
The proposed links to biodiversity, climate, or extinction remain controversial and are not assumed here; stellar engines serve only as physical motivation for a measurable acceleration.

Throughout this work, ``persistent'' or ``constant acceleration'' refers to the local sky-plane curvature represented by a \Gaia\ trend solution over the finite DR3 observing window.
The search is defined by this measured acceleration coefficient over the DR3 baseline.

Ordinary astrophysical and catalog effects can produce similar curvature.
A wide binary can mimic an approximately constant acceleration over a multi-year observing window.
Photocenter motion, crowding, variability, and scan-angle-dependent calibration residuals can also perturb a fitted astrometric solution \citep{Lindegren2021,GaiaDR3Validation2023}.
Moreover, the published \Gaia\ DR3 non-single-star acceleration solutions form a selected catalog subset rather than a random sample of all \Gaia\ stars \citep{GaiaDR3NSS2023,GaiaDR3Validation2023}.
A population-level technosignature test must therefore model ordinary accelerators and catalog systematics before translating the selected acceleration solutions into a constraint on nearby stars as a whole.

We define a quality-selected parent population of $406{,}984$ nearby stars, which provides the denominator for the population limits.
The likelihood is fitted to the $8{,}890$ published \Aseven\ solutions within this population.
We calibrate the astrometric-systematics tail from catalog controls, model long-period binary accelerations, and use the remaining allowed contribution to set threshold-dependent upper limits on stellar-engine-like transverse acceleration in the parent population.
Section~\ref{sec:data} defines the samples and observables, Section~\ref{sec:methods} describes the statistical model and population mapping, Section~\ref{sec:results} presents the limits, and Section~\ref{sec:discussion} discusses their physical interpretation.

\section{Samples and Observables}\label{sec:data}

\subsection{Gaia data and sample hierarchy}\label{subsec:datahierarchy}

The analysis uses public \Gaia\ DR3 products: source-level astrometry and photometry from \code{gaiadr3.gaia\_source}, NSS astrometric trend solutions from \code{gaiadr3.nss\_acceleration\_astro}, and astrophysical parameter estimates when a stellar mass is required for the power scale \citep{GaiaMission2016,GaiaDR3Summary2023,GaiaDR3NSS2023,GaiaApsis2023}.
Three sample definitions recur throughout the paper.
The \emph{parent sample} supplies the denominator for population limits; the published \emph{\Aseven\ sample} is fitted by the likelihood; and the \emph{mass subset} supplies the denominator for limits involving the photon-thrust-equivalent scale $P_0$.
Keeping them separate prevents a fitted fraction within the selected \Aseven\ catalog from being mistaken for an occurrence rate among all parent stars.
All reported population fractions refer to the explicitly defined parent sample or, for $P_0$, to the mass subset.

\subsection{Parent sample selection}\label{subsec:parentselection}

The parent sample is defined by the cuts in Table~\ref{tab:parentcuts}.
The cut $\varpi>5\ {\rm mas}$ corresponds, under the inverse-parallax conversion $d=1000/\varpi$ used here, to nominal distances $d<200\ {\rm pc}$, or about 650 light-years; the accompanying requirement $\varpi/\sigma_\varpi\ge20$ limits the formal fractional parallax uncertainty to at most 5\%.
The magnitude and visibility period cuts restrict the denominator to a regime with strong, well-sampled astrometry.
The duplicate source and image parameter cuts remove severe source matching and windowing artifacts that could otherwise dominate the astrometric tail.
We deliberately do not impose a hard RUWE veto in the parent definition: high RUWE is correlated with both genuine unresolved multiplicity and astrometric artifacts, and unresolved multiplicity is part of the astrophysical acceleration population that must be modeled rather than silently removed \citep{Lindegren2021,Kervella2022}.
RUWE is therefore retained for the systematics and searchable-fraction calculations instead of being used as a parent sample exclusion.

\begin{table*}[t]
\centering
\caption{Parent sample selection\label{tab:parentcuts}}
\footnotesize
\begin{tabular*}{\textwidth}{@{\extracolsep{\fill}}ll}
\toprule
Criterion & Requirement \\
\midrule
Astrometric solution & \code{astrometric\_params\_solved}$=31$ \\
Distance and parallax quality & $\varpi>5\ {\rm mas}$ and $\varpi/\sigma_\varpi\ge20$ \\
Magnitude range & $6<G<14$ \\
Observing coverage & \code{visibility\_periods\_used}$\ge10$ \\
Source matching & \code{duplicated\_source=false} \\
Image parameter diagnostics & \code{ipd\_frac\_multi\_peak}$\le2$ and \code{ipd\_frac\_odd\_win}$\le2$ \\
RUWE & no hard cut; retained for later modeling \\
\bottomrule
\end{tabular*}
\parbox{\textwidth}{\footnotesize\textit{Note.} Cuts are applied to \code{gaiadr3.gaia\_source} before intersection with the DR3 NSS trend solution table. The named quantities are standard \Gaia\ source-table astrometric and image-parameter quality diagnostics. They define the parent denominator $N_{\rm parent}=406{,}984$. Limits involving $P_0$ use the mass subset listed in Table~\ref{tab:samples}. RUWE is not used as an exclusion because it traces both astrometric artifacts and unresolved multiplicity.}
\end{table*}

\subsection{Acceleration solution sample}\label{subsec:channelsamples}

Before intersection with the parent sample, the full DR3 NSS trend solution table contains $338{,}215$ solutions: $246{,}947$ \Aseven\ solutions and $91{,}268$ \Anine\ solutions.
Intersecting this table with the parent sample by \code{source\_id} yields $15{,}936$ parent sample trend solutions, so $3.92\%$ of the parent sample has a published DR3 acceleration or acceleration plus jerk solution.
The solutions intersecting the parent sample split into $8{,}890$ \Aseven\ solutions and $7{,}046$ \Anine\ solutions (Table~\ref{tab:samples}).
The likelihood analysis is restricted to the \Aseven\ solutions because these are the \Gaia\ trend solutions most directly matched to an approximately constant transverse acceleration over the DR3 observing window.
The \Anine\ solutions are retained only as diagnostic context: they include jerk terms and therefore probe a different variable acceleration selection problem.
A statistically consistent joint analysis of \Aseven\ and \Anine\ would require an additional curvature/jerk selection model and an extra residual model term defined in the enlarged acceleration plus jerk space.
All primary limits in this paper are therefore explicitly conditional on the published \Aseven\ sample.

The mass-dependent power scale uses a separate denominator.
The parent sample contains $238{,}734$ stars with usable FLAME mass estimates.
Within the \Aseven\ likelihood sample, $4{,}405$ sources have usable mass estimates and can contribute to $P_0$ threshold exceedance counts.
No imputed mass is used to decide whether an \Aseven\ source exceeds a $P_0$ threshold; the $P_0$ numerator and denominator are both restricted to stars with mass estimates.

\begin{deluxetable*}{lcr}
\tablecaption{Sample hierarchy used for the population limits\label{tab:samples}}
\tablehead{\colhead{Quantity} & \colhead{Symbol} & \colhead{Value}}
\startdata
All-sky DR3 trend solutions in the full NSS table & $N_{\rm trend,all}$ & 338,215 \\
\Aseven\ solutions before parent intersection & $N_{\rm A7,all}$ & 246,947 \\
\Anine\ solutions before parent intersection & $N_{\rm A9,all}$ & 91,268 \\
Quality selected parent sample & $N_{\rm parent}$ & 406,984 \\
Trend solutions intersecting the parent sample & $N_{\rm trend\cap parent}$ & 15,936 \\
Fraction of parent stars with a published trend solution & $N_{\rm trend\cap parent}/N_{\rm parent}$ & 0.0392 \\
Constant acceleration solutions used in the likelihood & $N_{\rm A7}$ & 8,890 \\
Acceleration plus jerk solutions in the parent sample & $N_{\rm A9}$ & 7,046 \\
Parent stars with usable mass estimates for $P_0$ & $N_{\rm parent,mass}$ & 238,734 \\
\Aseven\ sources with usable masses for $P_0$ mapping & $N_{\rm A7,mass}$ & 4,405 \\
\enddata
\tablecomments{The all-sky entries summarize the published DR3 NSS trend solution catalog before applying the parent sample selection. The \Aseven\ subset is the sample fitted by the likelihood. The mass quantities are used only for the photon-thrust power scale and are not interchangeable with the full parent population denominator.}
\end{deluxetable*}

\subsection{Physical variables}\label{subsec:variables}

For a constant acceleration trend solution with fitted components on the sky $(\dot{\mu}_{\alpha^*},\dot{\mu}_{\delta})$, we define the angular acceleration amplitude
\begin{equation}
A=\left(\dot{\mu}_{\alpha^*}^2+\dot{\mu}_{\delta}^2\right)^{1/2},
\end{equation}
in \masyrtwo.
The corresponding transverse physical acceleration is
\begin{equation}
a_\perp = A\,d,
\end{equation}
with the conversion from \masyrtwo\ to rad s$^{-2}$ applied numerically and $d=1000/\varpi$ pc for parallax $\varpi$ in mas.
For stars with usable mass estimates, we also report the photon-thrust power scale
\begin{equation}
P_0 = M a_\perp c,
\end{equation}
in units of \lsun.
This is the perfectly collimated photon power that would supply the same force, not a physical model of an engineered system.
It should be read as a photon-thrust-equivalent momentum-flux scale: mass-loss or plasma-exhaust engines have a different relation between force, mass flux, exhaust speed, and mechanical power.
Because $P_0$ requires $M$, both the numerator and denominator of the $P_0$ limits use the parent mass subset; the relevant \Aseven\ residual weight sum is also restricted to the \Aseven\ sources with masses described in Table~\ref{tab:samples}.

\begin{figure*}[t]
\centering
\includegraphics[width=0.98\textwidth]{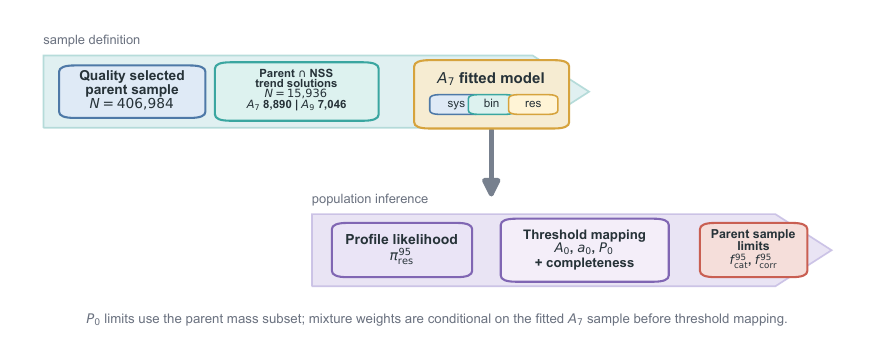}
\caption{Analysis workflow for the \Gaia\ DR3 astrometric stellar-engine limits. The upper lane defines the fitted sample, and the lower lane maps the profiled residual fraction to parent-sample limits. A quality-selected parent sample of $N=406{,}984$ stars is intersected with the published \Gaia\ DR3 NSS astrometric trend catalog, yielding $15{,}936$ trend solutions within the parent sample: $8{,}890$ \Aseven\ solutions and $7{,}046$ \Anine\ solutions. The primary likelihood uses only the constant-acceleration \Aseven\ sample. The mixture model includes empirical systematics, binary accelerations, and a residual persistent-acceleration component. A profile likelihood constrains the residual fraction in the fitted sample, which is then mapped to parent-sample threshold limits in $A_0$, $a_0$, and $P_0$ with an explicit searchable-fraction correction. Limits involving $P_0$ use the parent mass subset.}
\label{fig:workflow}
\end{figure*}

\section{Modeling the Published Acceleration Solutions}\label{sec:methods}

For each published \Aseven\ solution we use one observed quantity, the dimensionless \Gaia\ significance $s$.
We model its distribution as a mixture of empirical \Gaia\ systematics, long-period binary accelerations, and a residual persistent-acceleration component.
Here ``residual'' denotes the part of the significance distribution not absorbed by the adopted systematics and binary models.
Unmodeled binaries, photocenter effects, local systematics, missing quality information, or unusual accelerations can all contribute to this component.
The likelihood therefore constrains a population-level residual allowance after the leading conventional explanations have been included.
An active sky-plane momentum-transfer process could contribute to the local \Aseven\ curvature coefficient measured over the DR3 baseline.

\subsection{Catalog selection}\label{subsec:selection}

The published DR3 trend solutions form a selected catalog population.
We therefore fit the distribution of the published dimensionless significance statistic $s$ conditional on membership in the high-significance \Aseven\ sample.
The fixed threshold
\begin{equation}
S_{\rm cut}=20
\end{equation}
is adopted because \Gaia\ DR3 astrometric binary post processing required the acceleration solution significance $s$ to exceed 20 for both constant and variable acceleration models \citep{Halbwachs2023}.
The same processing also applied quality cuts, including $F_2<22$ for constant acceleration solutions, $F_2<25$ for variable acceleration solutions, and additional filtering related to parallax \citep{Halbwachs2023}.
Thus $S_{\rm cut}=20$ has a \Gaia\ DR3 origin, but it is not the complete NSS publication selection function; our likelihood conditions on the high-significance part of the published \Aseven\ sample rather than reconstructing the full catalog selection.
For any component density $p(s\mid\theta)$, the selected density is
\begin{equation}
p_{\rm sel}(s\mid\theta,S_{\rm cut})=
\frac{p(s\mid\theta)}{\int_{S_{\rm cut}}^{\infty}p(s'\mid\theta)\,ds'},\quad s\ge S_{\rm cut}.
\end{equation}
All likelihood terms below are normalized after this selection cut.
The fitted mixture weights describe only the published high-significance \Aseven\ sample.
Parent-population limits are obtained only after the residual-weight mapping and searchable-fraction calculation in Section~\ref{subsec:mapping}.

\subsection{Systematics tail calibration}\label{subsec:systematics}

The systematics model is calibrated directly from the DR3 trend solution table.
Sources are binned by ecliptic latitude and $G$ magnitude, capturing the dominant scan-law and centroiding-precision dependence while retaining enough objects per cell for stable tail estimation.
We attempted to use an external quasar reference-frame control as an inertial null sample, but its overlap with published NSS acceleration solutions was negligible; the final tail calibration therefore uses catalog controls drawn from the same published trend-solution population.
The control samples are defined using image parameter diagnostics and astrometric excess noise significance; RUWE is not used as a systematics flag because it is strongly entangled with real unresolved multiplicity.
A low-amplitude clean subset supplies a null-like control, and a low-amplitude subset enriched in systematics supplies a comparison control.
Both controls are split deterministically by source identifier into calibration and validation partitions.

\begin{figure*}[t]
\centering
\includegraphics[width=0.52\textwidth]{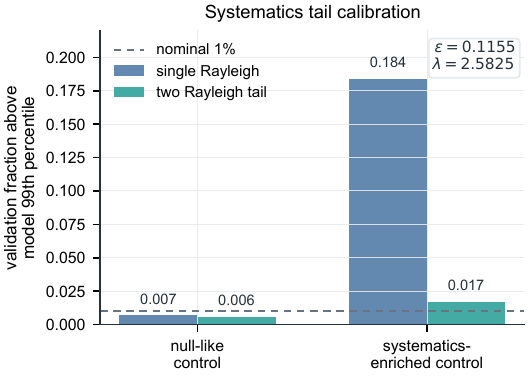}
\caption{Systematics-tail validation. If the tail model is calibrated, approximately 1\% of validation sources should exceed its predicted 99th percentile. A single-Rayleigh distribution is adequate for the null-like control but fails this test for the control enriched in systematics, producing too many high-significance exceedances. The adopted two-Rayleigh tail brings the exceedance rate substantially closer to the nominal level and reduces the risk that heavy-tailed \Gaia\ systematics are misallocated to the residual component.}
\label{fig:tailcal}
\end{figure*}

A Rayleigh tail is motivated when the two standardized acceleration components are approximately independent and Gaussian.
Validation shows that a single scale underpredicts extreme values in the control enriched in systematics.
We therefore adopt a two-Rayleigh mixture in which a broader minority component captures heavy-tailed \Gaia\ systematics and reduces their misallocation to the residual term.
For a selected Rayleigh tail the scale obeys
\begin{equation}
\mathbb{E}\left[s^2-S_{\rm cut}^2\mid s\ge S_{\rm cut}\right]=2\sigma^2.
\end{equation}
The calibration uses a robust median-based version of this estimator with hierarchical fallbacks from cell-level to marginal and global estimates.
The adopted selected tail model is
\begin{equation}
p_{\rm sys}(s)=(1-\epsilon)p_{\rm Ray}(s\mid\sigma_i)+
\epsilon p_{\rm Ray}(s\mid\lambda\sigma_i),
\end{equation}
with the same selection normalization applied to the mixture survival function.
The validation calibration gives
\begin{equation}
(\epsilon,\lambda)=(0.115475,2.582521).
\end{equation}
Here $\epsilon$ is the mixture weight of the broader systematics tail component, and $\lambda$ is the multiplicative scale factor of that broad component relative to the local Rayleigh scale.
As shown in Figure~\ref{fig:tailcal}, this model reduces the 99th percentile exceedance rate of the validation set enriched in systematics from 0.184 under a single-Rayleigh tail to 0.017, substantially closer to the nominal 0.01.

\subsection{Binary accelerators}\label{subsec:binary}

The binary component represents the dominant astrophysical confounder: companions whose orbital curvature is well approximated by constant acceleration over the DR3 astrometric baseline.
For each source we form an isotropic angular acceleration uncertainty proxy
\begin{equation}
\sigma_{A,{\rm iso}}=\left[\frac{\sigma_{A,1}^2+\sigma_{A,2}^2}{2}\right]^{1/2}
\end{equation}
from the two published acceleration component uncertainties.
For a binary draw, the projected physical acceleration on the sky is converted to angular units and then to the dimensionless expected signal amplitude, or noncentrality,
\begin{equation}
\delta=\kappa_{\rm eff}\frac{A}{\sigma_{A,{\rm iso}}}.
\end{equation}
The response factor $\kappa_{\rm eff}$ is fitted as a nuisance parameter rather than fixed from a detailed photocenter model.
This flexibility allows the binary term to represent a broad range of photocenter and acceleration-response effects, reducing the risk that unresolved binaries are assigned to the residual component.

The forward model samples the period $P$ from a lognormal distribution truncated to $P\in[P_{\min},10^6]$ yr, samples mass ratio from $p(q)\propto q^\gamma$ over $q\in[0.1,1]$, and marginalizes over viewing geometry.
These broad nuisance choices are intended to span the long-period companion tail rather than to impose a precise prior for field binaries \citep[e.g.,][]{Raghavan2010,MoeDiStefano2017}.
Circular orbit and thermal eccentricity prescriptions are both evaluated.
A wide binary can look almost like a constant acceleration because \Gaia\ observes only a short arc of the orbit.
The forward model therefore uses a finite-window response: it compares the observed trend to the constant-acceleration coefficient that would be recovered over the DR3 baseline, not to the instantaneous orbital second derivative.
We assume a uniform observing window of duration $T_{\rm obs}=33/12$ yr, representative of the DR3 astrometric baseline, and compute the response analytically for a circular harmonic displacement.
With $u=t/T_{\rm obs}\in[-1/2,1/2]$ and $x=2\pi T_{\rm obs}/P$, we fit $\cos(xu)$ by $c_0+c_1u+(1/2)c_2u^2$ and use
\begin{equation}
\eta(x)=\frac{|c_2|}{x^2},
\end{equation}
where
\begin{equation}
 c_2=\frac{120\sin(x/2)}{x}+\frac{720\cos(x/2)}{x^2}
      -\frac{1440\sin(x/2)}{x^3}.
\end{equation}
We restrict $\eta$ to $[0,1]$; for the assumed uniform circular window, the expression is exact and approaches $\eta\to1$ when $P\gg T_{\rm obs}$.
In thermal-eccentricity runs, orbital phase and instantaneous separation enter through the sampled projection, while the attenuation remains the circular finite-window response.
It is therefore an explicit approximation to epoch-level astrometry rather than a full epoch fit.
The selected binary likelihood is averaged over $K$ scrambled Sobol quasi Monte Carlo samples,
\begin{equation}
p_{{\rm bin},{\rm sel}}(s_i)\simeq \frac{1}{K}\sum_{k=1}^{K}
p_{\rm sel}\left(s_i\mid \delta_{ik}\right),
\end{equation}
where the conditional distribution at fixed noncentrality is Rice in the dimensionless significance variable.
The values of $K$ used in the sensitivity suite are numerical-integration choices, not physical hyperparameters.
The adopted $K=256$ case uses the largest Sobol sample size tested; Figure~\ref{fig:robustness} shows the sensitivity of the upper limit to $K=64,128,$ and 256.

\subsection{Residual component}\label{subsec:residual}

The residual component is where a stellar-engine-like acceleration would enter the model, but it can also absorb remaining conventional model mismatch.
It is deliberately broad and does not encode a specific engineering amplitude model.
At fixed noncentrality it also uses a selected Rice likelihood.
The residual noncentrality is marginalized over a log-uniform amplitude grid with
\begin{equation}
\delta\in[20,\delta_{\max}],\quad \delta_{\max}=\max(300,2s_{\max}),
\end{equation}
which gives $\delta_{\max}=461.97$ for the \Aseven\ sample.
The lower bound matches the fitted high-significance regime, while the upper bound extends beyond all observed values.
The grid assigns equal weight per logarithmic interval because the amplitude scale is unknown over several orders of magnitude; a linear grid would instead concentrate the residual kernel toward the largest amplitudes.
This choice defines the residual-component shape, and the reported limit remains conditional on it; the fitted parameter is the total residual mixture fraction after selection.
The mixture likelihood for the fitted sample is
\begin{equation}
\begin{split}
p_{\rm sel}(s_i)=&\;\pi_{\rm sys}p_{{\rm sys},{\rm sel}}(s_i)
+\pi_{\rm bin}p_{{\rm bin},{\rm sel}}(s_i)\\
&+\pirs p_{{\rm res},{\rm sel}}(s_i),
\end{split}
\end{equation}
with $\pi_{\rm sys}+\pi_{\rm bin}+\pirs=1$.

\subsection{Profile upper limit}\label{subsec:profile}

With the three likelihood components defined, the parameter to be limited is the residual mixture fraction $\pirs$.
We maximize the likelihood over mixture weights and binary nuisance parameters.
The upper limit on $\pirs$ is obtained by profiling the nuisance parameters at fixed $\pirs$ and finding the point where
\begin{equation}
\Delta\log\mathcal{L}=1.355,
\end{equation}
corresponding to the one-sided 95\% boundary-case likelihood-ratio calibration \citep{SelfLiang1987}.
We use multistart L-BFGS-B optimization.
At each fixed $\pirs$, the systematics and binary weights are constrained to satisfy $\pi_{\rm sys}+\pi_{\rm bin}=1-\pirs$, while the binary period location, period width, mass-ratio slope, and effective response scale are reoptimized.
The systematics tail calibration parameters $(\epsilon,\lambda)$ and the residual amplitude grid are held fixed for a given robustness configuration.
The profile crossing with $\log\mathcal{L}_{\rm max}-1.355$ is bracketed above the MLE and located by 20 bisection iterations; all upper limits on the adopted grid are bracketed.
We repeat the fit over a robustness suite varying the minimum binary period $P_{\min}$, the eccentricity prescription, the systematics tail prescription, and the number of Sobol samples.
We adopt the calibrated-tail, circular, $K=256$ result at $P_{\min}=10$ yr because it gives the largest upper limit among the evaluated $P_{\min}$ values within that configuration.
This is a conservative-envelope convention, not a model-selection claim.

\subsection{Mapping to parent sample limits}\label{subsec:mapping}

The fit to the published \Aseven\ sample does not by itself give a parent sample occurrence fraction.
We convert it in two steps: first by summing the residual weights above a chosen threshold, and then by dividing by the appropriate parent population denominator.
Let $r_{{\rm res},i}^{95}$ be the residual weight for object $i$ evaluated at the profiled 95\% upper limit parameter point.
For a threshold $x_0$ in $A$, $a_\perp$, or $P_0$, the expected residual count above threshold is
\begin{equation}
E_{\rm res}^{95}(>x_0)=\sum_{i\in \Aseven} r_{{\rm res},i}^{95}\,\mathbb{I}(x_i>x_0).
\end{equation}
The catalog-mapped parent-sample fraction limit, before the searchable-fraction adjustment, is
\begin{equation}
f_{\rm cat}^{95}(>x_0)=\frac{E_{\rm res}^{95}(>x_0)}{N_{\rm denom}},
\end{equation}
where $N_{\rm denom}=N_{\rm parent}$ for $A$ and $a_\perp$, and $N_{\rm denom}=N_{\rm parent,mass}$ for $P_0$.
Thus, $f_{\rm cat}^{95}$ expresses the residual weight above a threshold in the published \Aseven\ sample as a fraction of the appropriate parent-population denominator.

A separate completeness calculation estimates the searchable fraction $C(x_0)$ of the relevant parent population: the fraction for which a signal at threshold $x_0$ would exceed a chosen significance threshold.
This fraction is a detectability proxy under the adopted uncertainty model, not a reconstruction of the full NSS publication selection function.
The calculation uses a predictive model for the isotropic \Aseven\ angular-acceleration uncertainty $\sigma_{A,{\rm iso}}=[(\sigma_{A,1}^2+\sigma_{A,2}^2)/2]^{1/2}$, trained on published \Aseven\ solutions for which the covariance terms are defined.
The fitted target is $\log_{10}\sigma_{A,{\rm iso}}$.
The predictors are $G$, $G_{\rm BP}-G_{\rm RP}$, visibility periods, ecliptic latitude, RUWE, image parameter diagnostics, photometric excess factor, and Galactic latitude when available.
A histogram-gradient-boosting regressor is validated by holding out cells in $(G,{\rm ecliptic\ latitude})$ and is trained with density-ratio weights clipped to $0.2$--$5$ to reduce the mismatch between the published \Aseven\ sample and the parent population.
The validation diagnostics are ${\rm RMSE}=0.0709$ dex, median bias $0.0028$ dex, and 84th percentile true-minus-predicted residual $0.0649$ dex.
Before the searchable fraction is computed, the parent-sample uncertainty predictions are inflated by 0.0709 dex, a factor of 1.177.
Binned residual checks over magnitude, sky position, color, parallax proxy, and astrometric quality variables are used as validation diagnostics rather than as additional fitted corrections.
For a threshold in angular units, $C(A_0)$ is the fraction of parent stars with $A_0/\widehat{\sigma}_{A,{\rm iso},i}\ge S_{\min}$.
For $a_0$ and $P_0$, the same predicted angular uncertainty is converted to physical acceleration or photon-thrust power units using the parallax and, for $P_0$, the mass estimate; uncertainties in parallax and mass are not propagated into $C$.
The $f_{\rm corr}^{95}$ curves include this uncertainty inflation but do not marginalize over uncertainty in the predictive model.
Searchable fractions are reported for $S_{\min}=20$ and, for comparison, $S_{\min}=15$.
We quote a guarded searchable-fraction-adjusted limit
\begin{equation}
f_{\rm corr}^{95}(>x_0)=\frac{f_{\rm cat}^{95}(>x_0)}{\max[C(x_0),0.05]}.
\end{equation}
The 0.05 floor prevents the correction from diverging where the parent sample is effectively unsearchable.
Values set by the searchable-fraction floor do not provide strong occurrence constraints in poorly searchable regimes; they are extrapolated sensitivity indicators rather than precise population limits.

\section{Results}\label{sec:results}

We first report the likelihood result for the published high-significance \Aseven\ sample and then translate it into acceleration limits for the parent population.
The fitted-sample fraction is an intermediate quantity; the parent-population curves are the physical result, with their interpretation set by the searchable fraction.

\subsection{Fitted sample limit}\label{subsec:fit}

The adopted fit contains $8{,}890$ \Aseven\ solutions above $S_{\rm cut}=20$.
At the maximum-likelihood point the mixture weights are
\begin{equation}
(\pi_{\rm sys},\pi_{\rm bin},\pirs)=(0.3317,0.6394,0.0288).
\end{equation}
The adopted profiled upper limit is
\begin{equation}
\pirs^{95}=0.035087.
\end{equation}
This number is a residual fraction within the published \Aseven\ sample, conditional on the adopted systematics and binary models.
It enters the parent-population calculation only through the threshold mapping described above.
Table~\ref{tab:fit} summarizes the adopted configuration and primary fitted-sample quantities.

\begin{deluxetable*}{lcc}
\tablecaption{Adopted likelihood configuration and fitted sample result\label{tab:fit}}
\tablehead{\colhead{Quantity} & \colhead{Symbol} & \colhead{Value}}
\startdata
Fitted sample threshold & $S_{\rm cut}$ & 20 \\
Likelihood sample size & $N_{\rm A7}$ & 8,890 \\
Systematics broad-tail fraction & $\epsilon$ & 0.115475 \\
Systematics tail scale multiplier & $\lambda$ & 2.582521 \\
Minimum binary period in adopted configuration & $P_{\min}$ & 10 yr \\
Sobol quasi Monte Carlo samples & $K$ & 256 \\
Fitted sample systematics fraction at MLE & $\pi_{\rm sys}$ & 0.3317 \\
Fitted sample binary fraction at MLE & $\pi_{\rm bin}$ & 0.6394 \\
Fitted sample residual fraction at MLE & $\pirs$ & 0.0288 \\
One-sided 95\% fitted sample residual limit & $\pirs^{95}$ & 0.0351 \\
\enddata
\tablecomments{Here $S_{\rm cut}$ is the catalog significance threshold used to define the fitted \Aseven\ sample. The residual fraction is conditional on that published sample. Parent sample limits are obtained only after mapping the residual weights and applying the searchable-fraction adjustment.}
\end{deluxetable*}
\subsection{Robustness checks}\label{subsec:robustness}

\begin{figure*}[t]
\centering
\includegraphics[width=0.90\textwidth]{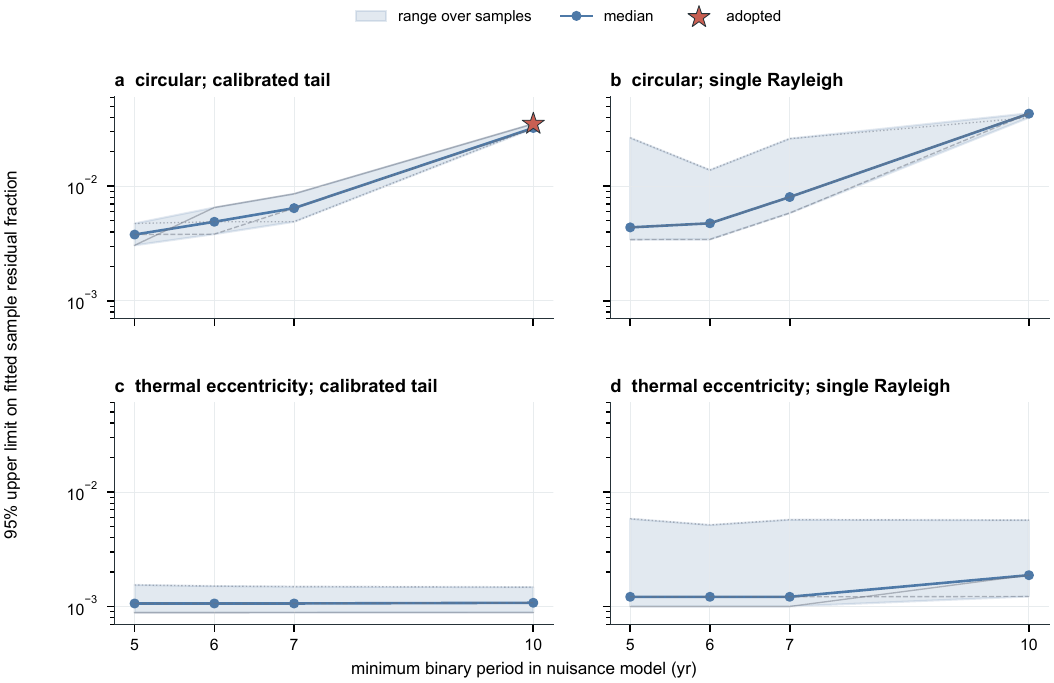}
\caption{Sensitivity of the one-sided 95\% upper limit on the fitted-sample residual fraction. Each panel varies the minimum binary period. The shaded region spans the three Sobol sample sizes, and the solid line shows their median; gray guide lines show individual sample sizes. The star marks the adopted conservative-envelope result in the calibrated-tail circular panel.}
\label{fig:robustness}
\end{figure*}

The robustness checks contain 48 fits spanning $P_{\min}=5,6,7,$ and 10 yr, circular and thermal eccentricity prescriptions, single-Rayleigh and calibrated two-Rayleigh systematics tails, and $K=64,128,$ and 256 Sobol samples.
These checks separate physical nuisance assumptions from numerical-integration choices: $P_{\min}$ and eccentricity change the binary tail, the Rayleigh prescription changes the systematics tail, and $K$ measures sensitivity to the Sobol sample size.
Figure~\ref{fig:robustness} shows the resulting 95\% upper limits.
Within this forward model, the thermal-eccentricity prescription produces smaller residual limits because eccentric companions populate the high-significance tail more efficiently.
The single-Rayleigh systematics test can increase the limit because its narrower tail allocates more probability to the other components.
Among the calibrated-tail circular fits evaluated at $K=256$, the $P_{\min}=10$ yr configuration gives the largest upper limit and is therefore adopted as the conservative envelope.
The corresponding single-Rayleigh circular value is $0.0434$, but that configuration is retained only as a sensitivity test because it fails the empirical tail calibration.

\subsection{Parent sample limits}\label{subsec:limits}

The parent-sample curves connect the fitted result to physical acceleration thresholds.
They map residual weights from the published \Aseven\ sample to the defined parent population above each chosen threshold.
We report three scales: angular acceleration $A_0$, transverse physical acceleration $a_0$, and the photon-thrust-equivalent scale $P_0$.
Figure~\ref{fig:limits} shows two curves for each scale.
The catalog-mapped curve counts residual weight in the published \Aseven\ sample and divides by the appropriate parent-population denominator.
The adjusted curve then accounts for the fraction of parent stars whose threshold signal would exceed the adopted significance criterion under the uncertainty proxy.
It is most directly interpretable where this searchable fraction is high; at low searchable fraction, it is a sensitivity indicator rather than a strong occurrence constraint.
For the angular acceleration scale, the catalog-mapped limit is $7.20\times10^{-4}$ at $A_0=1.00\ {\rm mas\ yr^{-2}}$, $6.00\times10^{-4}$ at $A_0=2.00\ {\rm mas\ yr^{-2}}$, and $2.47\times10^{-5}$ at $A_0=10.0\ {\rm mas\ yr^{-2}}$.
The physical acceleration catalog-mapped limit is $7.28\times10^{-4}$ at $a_0=1.08\times10^{-5}\ {\rm m\ s^{-2}}$, $3.66\times10^{-4}$ at $a_0=9.68\times10^{-5}\ {\rm m\ s^{-2}}$, and $1.21\times10^{-5}$ at $a_0=2.60\times10^{-4}\ {\rm m\ s^{-2}}$.
At the lowest $a_0$ threshold, however, the $S_{\min}=20$ searchable fraction is only 0.070, so the adjusted limit is much weaker, $1.04\times10^{-2}$.
Once the searchable fraction approaches unity, the catalog-mapped and adjusted curves coincide.

The same pattern appears in $P_0$.
The representative $P_0$ thresholds start at $P_0=1.39\times10^7\ \lsun$, where $f_{\rm cat}^{95}=5.96\times10^{-4}$ and $C_{20}=0.0447<0.05$.
This point is below the adopted searchable-fraction floor, so its $S_{\min}=20$ correction is set by that floor and should be read only as a low-threshold edge case.
At the higher representative threshold $P_0=2.69\times10^7\ \lsun$, $C_{20}=0.272$ and $f_{\rm cat}^{95}=5.80\times10^{-4}$.
At $P_0=1.94\times10^8\ \lsun$, where the searchable fraction is 0.996, the adjusted and catalog-mapped limits are both approximately $2.1\times10^{-4}$.

\onecolumngrid
\begin{center}
\refstepcounter{figure}\label{fig:limits}
\includegraphics[width=0.80\textwidth]{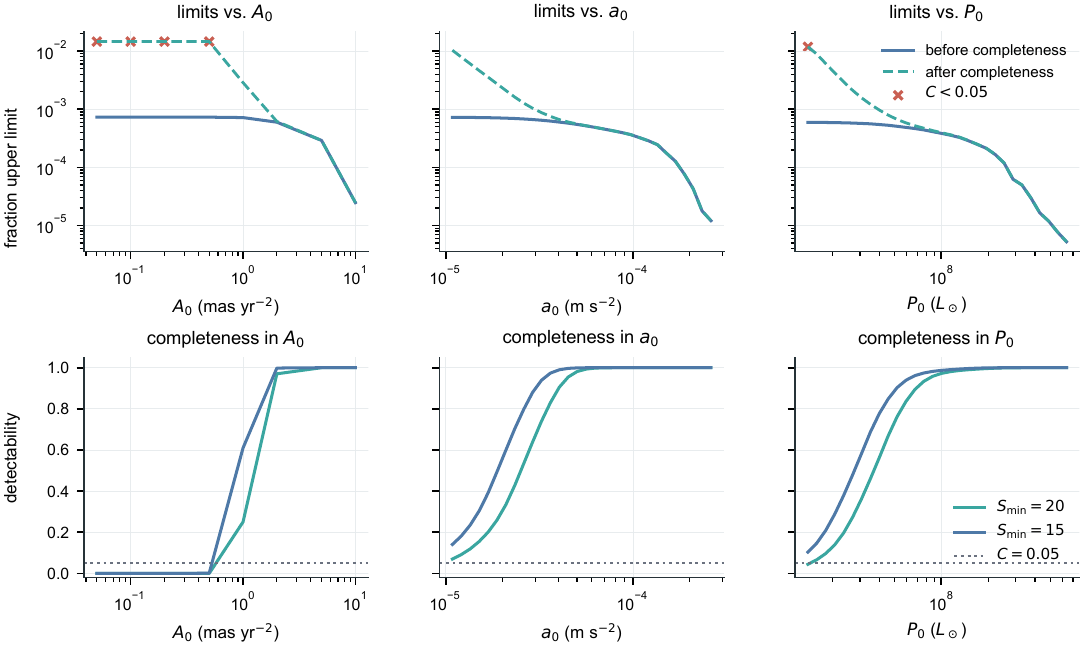}
\parbox{0.84\textwidth}{\footnotesize\textbf{Figure~\thefigure.} Parent-sample upper limits and searchable fractions. Top row: 95\% limits before and after the searchable-fraction correction. Crosses mark thresholds where the $S_{\min}=20$ searchable fraction is below 0.05, so the adjusted curve uses the floor and should be read only as a sensitivity indicator. Bottom row: searchable fractions used for the correction. High-threshold regions with $C\simeq1$ are the most directly interpretable.}
\end{center}
\twocolumngrid
\hypertarget{section.5}{}\phantomsection

\section{Discussion}\label{sec:discussion}

\subsection{Physical meaning of the acceleration scale}\label{subsec:accelvelocity}

A stellar engine transfers momentum to a star, so it can produce a dynamical acceleration signature while operating.
For scale, a constant acceleration of $10^{-4}\ {\rm m\ s^{-2}}$ would produce a velocity change of approximately $0.01c$ in $10^3$ yr and $0.1c$ in $10^4$ yr.
These timescales show how rapidly such a large acceleration would alter a star's motion.
The analysis therefore constrains the fraction of the defined parent population allowed to show this high-acceleration signature during the DR3 epoch; inferring the duration or lifetime frequency of such phases would require an additional population model.

\subsection{Power scale and engine mechanism}\label{subsec:p0discussion}

The $P_0=M a_\perp c$ axis is a photon-thrust-equivalent scale: the luminosity of a perfectly collimated photon stream that supplies the same force. It puts accelerations of stars with different masses onto a common momentum-flux scale. Mass-loss engines, plasma exhaust concepts, and binary-engine designs can have different energy and momentum budgets, so the most direct observables remain $A_0$ and $a_0$, with $P_0$ serving as an interpretable force normalization.

\subsection{False positives and false negatives}\label{subsec:falseposneg}

The main false-positive channels in this astrometric search are unresolved binaries, luminous companions, photocenter motion, crowding, scan-law-dependent calibration residuals, and incompletely captured image-parameter pathologies.
The empirical systematics and binary terms allocate these conventional explanations before the residual component is bounded.
False negatives include coasting systems whose engine phase has ended, accelerations below the \Gaia\ DR3 trend-solution threshold, mostly radial accelerations, and signals absorbed by the flexible binary or systematics components.

\subsection{Limitations}\label{subsec:uncertainties}

The largest systematic uncertainty is the tail behavior of the published acceleration significance statistic.
The validation controls strongly favor a two-Rayleigh tail over a single-Rayleigh tail for the population enriched in systematics, and that calibration is therefore used in the adopted model.
Nevertheless, the control construction is based on catalog quantities and cannot fully reproduce all local scan-law, crowding, and color-dependent effects.
Future releases with longer time baselines and richer calibration products should make this component more empirical and less parametric.

The binary model is flexible but still simplified.
It uses broad long-period distributions, a fitted response factor, and discrete eccentricity prescriptions; it does not attempt detailed population synthesis or photocenter modeling for each source.
Allowing the binary component this flexibility reduces false attribution to the residual term, but it can also absorb genuine persistent accelerations.
The result therefore remains conditional on the adopted component shapes and nuisance-parameter ranges.

The searchable-fraction adjustment is another important limitation.
At thresholds where the searchable fraction is near unity, the adjustment is small and the parent-sample limits are most directly interpretable.
Where the searchable fraction is small, values set or strongly affected by the floor indicate sensitivity rather than precise occurrence constraints.
This is particularly important for the lowest angular acceleration thresholds and for the lowest $P_0$ thresholds.

\subsection{Future extensions}\label{subsec:prospects}

Several extensions would strengthen the analysis.
A joint model of \Aseven\ and \Anine\ solutions would use more of the trend solution information and could separate persistent accelerations from curvature changes over the observing window.
A source-level binary prior, informed by color, luminosity, resolved companions, and external radial velocities, would reduce the residual flexibility now absorbed by the binary component.
Finally, independent astrometry from future \Gaia\ releases, Hipparcos--\Gaia\ proper-motion anomalies, or ground-based high-angular-resolution imaging can turn the current population accounting into source-level vetting \citep{Brandt2021,Kervella2022}.

\section{Conclusions}\label{sec:conclusions}

For the quality-selected parent population of $406{,}984$ \Gaia\ DR3 stars with nominal distances below 200 pc, our clearest constraint applies above a transverse-acceleration threshold of $a_0=2.60\times10^{-4}\ {\rm m\ s^{-2}}$, where the modeled searchable fraction is essentially unity.
The one-sided 95\% model-conditional upper limit on the fraction of this population allowed to show such accelerations is $1.21\times10^{-5}$---about 12 per million stars in the parent population.

We obtain this constraint by fitting the $8{,}890$ published constant-acceleration (\Aseven) solutions with a mixture model that accounts for long-period binaries and catalog systematics, and then mapping the remaining allowed contribution back to the parent population.
At the headline threshold, the limit corresponds to a summed residual-component weight of about five star-equivalents across the full parent sample; this is a population-level statistical equivalent, not a count of identified or detected systems.
At lower acceleration thresholds, the modeled searchable fraction declines, so the corresponding bounds provide weaker population constraints and mainly describe the reach of the search.
This work therefore turns \Gaia\ astrometric curvature into a quantitative population test of very strong, stellar-engine-like transverse acceleration, while longer observing baselines and future \Gaia\ releases can extend the same approach to weaker accelerations.

\begin{acknowledgments}
This work was supported by the National Key R\&D Program of China (No.\ 2024YFA1611804), the Shandong Provincial Natural Science Foundation, China (No.ZR2019MA059),
the China Manned Space Program (CMS-CSST2025-A01), the National SKA Program of China under Grant No. 2025SKA0120104, the Shandong Provincial Natural Science Foundation (ZR2024QA180),
and the Scientific Research Fund of Dezhou University (4022504019).
This work uses data from the European Space Agency mission \Gaia, processed by the \Gaia\ Data Processing and Analysis Consortium.
Funding for the DPAC has been provided by national institutions, in particular the institutions participating in the \Gaia\ Multilateral Agreement.

\end{acknowledgments}

\bibliographystyle{aasjournalv7}
\bibliography{references}

\end{document}